# Wireless LAN techniques for mobile instrument control

Jörn Beckmann, Jörg Pulz

Technische Universität München
ZWE FRM-II
D-85747 Garching, Germany

A choice of wireless technologies is available today. Not all of them are suitable to be used in a instrument control network at a neutron scattering facility. This paper will start with an overview of the wireless communication protocols IrDA, IEEE 802.11 WLAN, Bluetooth and protocols used by cellular phones. Afterwards some thoughts on network layout and the results of tests performed with 802.11b and Bluetooth equipment are presented.

## 1 Introduction to existing wireless communication techniques

Today's instrument control software is mainly based on a client-server architecture, requiring a network infrastructure to operate flexible instrument setups, are often disturbed by network cabling. Another trend is to facilitate pocket-size computers, so called PDAs, for all sort of work. To use such a device in instrument control duty, it has to be connected to the network. Both fields of application strongly demand for wireless networking. Several techniques have been developed during the last years. Below we will discuss major wireless techniques, which are able to support networking applications.

### 1.1 IrDA
Nearly all current laptops bear an interface compatible with the standard of the Infrared Data Association (IrDA) allowing ad-hoc communication in changing environments. The standard foresees point-to-point connections in a range up to 1 m and a maximal angle of 15 degrees. Connection speed lies between 9.6 kbit/sec and 16 Mbit/sec depending on the protocol used. As infrared light is used as the means of communication a direct line of sight is required. Application specific protocols are IrCOMM (emulating of serial and parallel interfaces), IrLPT (wireless printer connection), IrOBEX (exchange of structured data, as used by a Palm Pilot) and IrLAN (offering network connections). IrDa compatible hardware is widespread and supported by major operating Systems like Windows and Linux (included since 2.2 Kernel versions). There are no methods for authentication or encrypted data transfer defined in the standard. Because of its limited range and as it only supports point-to-point connections, IrDA is not the right choice for robust and reliable wireless instrument control.

### 1.2  Wireless LAN (IEEE 802.11)

The usage of radio frequencies in data communication was first regulated in 1997 by the IEEE standard 802.11, which allows data rates of 1 or 2 Mbit/sec and offers a range of 300 m under direct line of sight and up to 30 m within buildings. Wireless networks following the IEEE 802.11 standard operate on 79 frequency slots around 2.4 GHz either by means of Frequency Hopping Spread Spectrum (FHSS) or Direct Sequence Spread Spectrum (DSSS). In the ad-hoc mode only point-to-point connections are possible while the infrastructure mode allows the creation of structured networks by means of access points. The only security feature of the 802.11 standard is the network name (SSID), which has to set to the same value on all devices who want to communicate with each other. All frequencies are separated into 13 channels but these channels overlap in a way that only 4 different WLANs per installation are possible.

An extension to a data rate of up to 11 Mbit/sec could be found in the 802.11b standard. It also operates in the 2.4 GHz range. A further extension from 802.11b is Wire Equivalent Privacy (WEP). WEP uses the RC4 stream cipher to encrypt transmitted data. First a 32-bit Integrity Check Value (ICV) is calculated and placed behind the actual message. This new message is encrypted by XORing it with the constant stream key (WEP40 uses 40 bits, WEP128 uses 104 bits) and a variable Initialization Vector (IV). The IV changes for each transmitted package. As the receiver needs to know the actual value of IV, it is placed in plaintext in front of the transmitted package. Together with the secret WEP key the receiver can now decrypt the message. The idea behind this is, that a retransmission of the same message should not use the same encrypted package as this would allow an attacker to figure out the key. Unfortunately the IV is only 24 bits long and so the same IV is regularly reused. An attacker needs to sniff 1000 to 4000 "interesting" packages to recalculate the secret key. For WEP40 one out of 1000 transmitted packages is "interesting" for the attacker, in the case of WEP128 about one package out of 10000 is interesting. In a medium used WLAN it will take three or four days of sniffing to break an WEP key. Therefore WEP is believed to be unsafe today. A second problem with WEP is, that equipment from different vendors is often unable to communicate while WEP is active.

Meanwhile WLAN is able to transmit up to 54 Mbit/sec. This is provided by the IEEE 802.11a standard. It uses the Orthogonal Frequency Division Multiplexing (OFDM) and works at 5 GHz. Therefore equipment that fulfills the 802.11a standard is unable to communicate with equipment from 802.11b. The second problem is, that in Europe HiperLAN/2 works also in the 5 GHz region and IEEE 802.11a was approved just some month ago by ETSI. A standard which will allow to transmit 54 Mbit/sec in the

2.4 GHz region is IEEE 802.11g. This new standard is backwards compatible with the older WLAN standards but is still under development. IEEE 802.11 and IEEE802.11b wireless LAN are supported by all major operating systems.

## 1.3 Bluetooth

Originally designed to replace serial cables, Bluetooth meanwhile allows communication between lots of different devices. The Bluetooth standard is developed by the Bluetooth Special Interest Group. Only members of that consortium may get their equipment certified. Bluetooth also works in the 2.4 GHz range and therefore Bluetooth and WLAN disturb each other. Bluetooth either transmits 433 kbit/sec symmetrically or 721 kbit/sec upstream and 58 kbit/sec downstream. The range is up to 10 m for Micro-Bluetooth, which is commonly used. The Bluetooth specification contains several profiles which supply a certain service. The Headset Profile for example supplies wireless connections between a cellular phone and a headset. The Serial Cable Emulation RFCOMM is designed to replace IrDA and the LAA Access Profile offers network connections via PPP.

The strongest point in favor of Bluetooth is security. The standard contains symmetric encryption, Authentication and Authorization. During Authentication the devices identify each other by means of a 128 bit one-time Initialization key. This key is made up of a PIN, device address, master clock and a random number. With that key a 128 bit Link Key is established which will be used to encrypt further communication. Creation of a link between devices either has to be allowed explicitly or the devices might be paired which will allow an automatic reconnection next time the devices find each other.

Bluetooth is supported by Linux since Kernel version 2.4.6 by means of the blueZ package. Windows does not offer generic Bluetooth support in the moment, device manufacturers have to provide the necessary drivers and software.

## 1.4 GSM/HSCSD/GPRS/UMTS

Finally protocols in use by cellular phones could be used to transmit data in wireless instrument control applications. The GSM standard allows only communication with 9.4 kbit/sec. This is extended up to 56 kbit/sec by either HSCSD or GPRS. HSCSD bundles GSM channels while GPRS provides packet oriented communication and allows better bandwidth usage. UMTS finally allows communication in the Megabit region but it is only used in pilot installations for the moment. All four protocols have the disadvantage that they operate in public networks and are charged by telecommunication companies. They all require contact to base stations and are therefore not usable in the neighborhood of heavy shielding.

# 2 Setup of instrument networks

At FRM-II instrument control software is based on the Taco system from the ESRF. The system heavily uses RPC calls. To prevent accidental submission of commands to wrong instruments, each instrument operates an own subnet. The subnets are separated by VLANs and connection between VLANs is checked by access control lists. Such a strict separation of networks is also desired in wireless instrument control networks. With Bluetooth this is easy to achieve by the pairing mechanism. For IEEE 802.11 networks different SSID and WEP keys for each instrument should prevent at least accidental interference. But a malfunctioning 802.11 sender will affect all networks in range as bandwidth is shared between stations even with different network names. Due to the channel setup only four parallel networks are possible, which is not enough as the experimental hall at FRM-II houses 10 experiments. As each independent network requires its own access point this increases costs dramatically. For this reason all wireless clients have to share one wireless network. The access point has to divide the incoming traffic from mobile devices and routes it to the individual instrument networks. IP or MAC address are not enough to separate traffic as they are both set on the mobile device (unfortunately most wireless card drivers allow to modify the MAC address). A good way is the usage of IPsec in connection with authentication to a RADIUS server. This will require additional software on the client and the servers in the instrument network and is not possible in certain situations. Further research has to be done in order to find a suitable and easy to use solution.

## 3      Test Results

Before much effort is put into wireless instrumentation control, it needs to be checked whether wireless communication is possible at the facility. We used a Compaq iPAQ 3570 and an Anycom BT Access Point to check Bluetooth communication. In normal office environment both device paired without problems and reached the 10 m distance between Access Point and mobile client. After the connection was lost however the reconnect did not work at all times. From time to time Access Point and mobile client had to be reset to build up a new connection. In the experimental hall the situation became even worse as the maximal distance between client and access point was reduced to 6 m and the reconnect failed nearly always.

Tests of IEEE 802.11b equipment were performed with a Lucent Orinoco Gold PCMCIA card and an Avaya Access point containing the same card. As a second mobile client a Sharp SL-5000 (Zaurus) with a D-Link DCF-650W WLAN CF-Card was used. The Sharp SL-5000 runs with preinstalled Linux and recognized the WLAN CF-Card instantly. One access point was enough to cover the whole experimental hall (aprox. 30 m x 30 m) when using an Orinoco Gold, even with the reactor polygon between Access point and client. The SL-5000 had slight connection problems behind the polygon and in corners of instrument shielding. A second Access point should overcome these problems. From outside the experimental hall no connection was possible to the Access Point inside.

A second test was done in the Guide Hall (50 m x 24 m), which is half empty in the moment. Also in the Guide Hall one Access Point was sufficient. Outside the Guide Hall building no connection to the Access Point inside was possible. From nearby offices with a window to the Guide Hall a connection with reduced data rate was possible. As access to that offices and the visitors windows is not restricted extra measurements have to be taken before a wireless network can be established here. Otherwise WEP keys might be sniffed from either the offices or the visitors window.

## 4      Conclusion and further planning

For the moment only IEEE 802.11b seems to be suitable for the setup of a wireless instrument control network. However there are security issues open which have to be dealt with before such a network can be commissioned. This will also allow 802.11a or 802.11g to be established on the market. Both are interesting for speed reasons. But as 802.11a is not backward compatible with 802.11, one has to either implement an additional 802.11(b) wireless network or wait for 802.11g.